\documentstyle[aas2pp4]{article}
\def\PsfigVersion{1.10}
\def\setDriver{\DvipsDriver} 
\ifx\undefined\psfig\else\endinput\fi
\let\LaTeXAtSign=\@
\let\@=\relax
\edef\psfigRestoreAt{\catcode`\@=\number\catcode`@\relax}
\catcode`\@=11\relax
\newwrite\@unused
\def\ps@typeout#1{{\let\protect\string\immediate\write\@unused{#1}}}

\def\DvipsDriver{
	\ps@typeout{psfig/tex \PsfigVersion -dvips}
\def\PsfigSpecials{\DvipsSpecials} 	\def\ps@dir{/}
\def\ps@predir{} }
\def\OzTeXDriver{
	\ps@typeout{psfig/tex \PsfigVersion -oztex}
	\def\PsfigSpecials{\OzTeXSpecials}
	\def\ps@dir{:}
	\def\ps@predir{:}
	\catcode`\^^J=5
}

\def\figurepath{./:}

\def\DoPaths#1{\expandafter\EachPath#1\stoplist}
\def\leer{}
\def\EachPath#1:#2\stoplist{
  \ExistsFile{#1}{\SearchedFile}
  \ifx#2\leer
  \else
    \expandafter\EachPath#2\stoplist
  \fi}
\def\ps@dir{/}
\def\ExistsFile#1#2{%
   \openin1=\ps@predir#1\ps@dir#2
   \ifeof1
       \closein1
   \else
       \closein1
        \ifx\ps@founddir\leer
           \edef\ps@founddir{#1}
        \fi
   \fi}
\def\get@dir#1{%
  \def\ps@founddir{}
  \def\SearchedFile{#1}
  \DoPaths\figurepath
}

\def\@nnil{\@nil}
\def\@empty{}
\def\@psdonoop#1\@@#2#3{}
\def\@psdo#1:=#2\do#3{\edef\@psdotmp{#2}\ifx\@psdotmp\@empty \else
    \expandafter\@psdoloop#2,\@nil,\@nil\@@#1{#3}\fi}
\def\@psdoloop#1,#2,#3\@@#4#5{\def#4{#1}\ifx #4\@nnil \else
       #5\def#4{#2}\ifx #4\@nnil \else#5\@ipsdoloop #3\@@#4{#5}\fi\fi}
\def\@ipsdoloop#1,#2\@@#3#4{\def#3{#1}\ifx #3\@nnil 
       \let\@nextwhile=\@psdonoop \else
      #4\relax\let\@nextwhile=\@ipsdoloop\fi\@nextwhile#2\@@#3{#4}}
\def\@tpsdo#1:=#2\do#3{\xdef\@psdotmp{#2}\ifx\@psdotmp\@empty \else
    \@tpsdoloop#2\@nil\@nil\@@#1{#3}\fi}
\def\@tpsdoloop#1#2\@@#3#4{\def#3{#1}\ifx #3\@nnil 
       \let\@nextwhile=\@psdonoop \else
      #4\relax\let\@nextwhile=\@tpsdoloop\fi\@nextwhile#2\@@#3{#4}}
\ifx\undefined\fbox
\newdimen\fboxrule
\newdimen\fboxsep
\newdimen\ps@tempdima
\newbox\ps@tempboxa
\long\def\fbox#1{\leavevmode\setbox\ps@tempboxa\hbox{#1}\ps@tempdima\fboxrule
    \advance\ps@tempdima \fboxsep \advance\ps@tempdima \dp\ps@tempboxa
   \hbox{\lower \ps@tempdima\hbox
  {\vbox{\hrule height \fboxrule
          \hbox{\vrule width \fboxrule \hskip\fboxsep
          \vbox{\vskip\fboxsep \box\ps@tempboxa\vskip\fboxsep}\hskip 
                 \fboxsep\vrule width \fboxrule}
                 \hrule height \fboxrule}}}}
\fi
\newread\ps@stream
\newif\ifnot@eof       
\newif\if@noisy        
\newif\if@atend        
\newif\if@psfile       
{\catcode`\%=12\global\gdef\epsf@start{
\def\epsf@PS{PS}
\def\epsf@getbb#1{%
\openin\ps@stream=\ps@predir#1
\ifeof\ps@stream\ps@typeout{Error, File #1 not found}\else
   {\not@eoftrue \chardef\other=12
    \def\do##1{\catcode`##1=\other}\dospecials \catcode`\ =10
    \loop
       \if@psfile
	  \read\ps@stream to \epsf@fileline
       \else{
	  \obeyspaces
          \read\ps@stream to \epsf@tmp\global\let\epsf@fileline\epsf@tmp}
       \fi
       \ifeof\ps@stream\not@eoffalse\else
       \if@psfile\else
       \expandafter\epsf@test\epsf@fileline:. \\%
       \fi
          \expandafter\epsf@aux\epsf@fileline:. \\%
       \fi
   \ifnot@eof\repeat
   }\closein\ps@stream\fi}%
\long\def\epsf@test#1#2#3:#4\\{\def\epsf@testit{#1#2}
			\ifx\epsf@testit\epsf@start\else
\ps@typeout{Warning! File does not start with `\epsf@start'.  It may not be a PostScript file.}
			\fi
			\@psfiletrue} 
{\catcode`\%=12\global\let\epsf@percent=
\long\def\epsf@aux#1#2:#3\\{\ifx#1\epsf@percent
   \def\epsf@testit{#2}\ifx\epsf@testit\epsf@bblit
	\@atendfalse
        \epsf@atend #3 . \\%
	\if@atend	
	   \if@verbose{
		\ps@typeout{psfig: found `(atend)'; continuing search}
	   }\fi
        \else
        \epsf@grab #3 . . . \\%
        \not@eoffalse
        \global\no@bbfalse
        \fi
   \fi\fi}%
\def\epsf@grab #1 #2 #3 #4 #5\\{%
   \global\def\epsf@llx{#1}\ifx\epsf@llx\empty
      \epsf@grab #2 #3 #4 #5 .\\\else
   \global\def\epsf@lly{#2}%
   \global\def\epsf@urx{#3}\global\def\epsf@ury{#4}\fi}%
\def\epsf@atendlit{(atend)} 
\def\epsf@atend #1 #2 #3\\{%
   \def\epsf@tmp{#1}\ifx\epsf@tmp\empty
      \epsf@atend #2 #3 .\\\else
   \ifx\epsf@tmp\epsf@atendlit\@atendtrue\fi\fi}

\chardef\psletter = 11 
\chardef\other = 12

\newif \ifdebug 
\newif\ifc@mpute 
\c@mputetrue 

\let\then = \relax
\def\r@dian{pt }
\let\r@dians = \r@dian
\let\dimensionless@nit = \r@dian
\let\dimensionless@nits = \dimensionless@nit
\def\internal@nit{sp }
\let\internal@nits = \internal@nit
\newif\ifstillc@nverging
\def \Mess@ge #1{\ifdebug \then \message {#1} \fi}

{ 
	\catcode `\@ = \psletter
	\gdef \nodimen {\expandafter \n@dimen \the \dimen}
	\gdef \term #1 #2 #3%
	       {\edef \t@ {\the #1}
		\edef \t@@ {\expandafter \n@dimen \the #2\r@dian}%
		\t@rm {\t@} {\t@@} {#3}%
	       }
	\gdef \t@rm #1 #2 #3%
	       {{%
		\count 0 = 0
		\dimen 0 = 1 \dimensionless@nit
		\dimen 2 = #2\relax
		\Mess@ge {Calculating term #1 of \nodimen 2}%
		\loop
		\ifnum	\count 0 < #1
		\then	\advance \count 0 by 1
			\Mess@ge {Iteration \the \count 0 \space}%
			\Multiply \dimen 0 by {\dimen 2}%
			\Mess@ge {After multiplication, term = \nodimen 0}%
			\Divide \dimen 0 by {\count 0}%
			\Mess@ge {After division, term = \nodimen 0}%
		\repeat
		\Mess@ge {Final value for term #1 of 
				\nodimen 2 \space is \nodimen 0}%
		\xdef \Term {#3 = \nodimen 0 \r@dians}%
		\aftergroup \Term
	       }}
	\catcode `\p = \other
	\catcode `\t = \other
	\gdef \n@dimen #1pt{#1} 
}

\def \Divide #1by #2{\divide #1 by #2} 

\def \Multiply #1by #2
       {{
	\count 0 = #1\relax
	\count 2 = #2\relax
	\count 4 = 65536
	\Mess@ge {Before scaling, count 0 = \the \count 0 \space and
			count 2 = \the \count 2}%
	\ifnum	\count 0 > 32767 
	\then	\divide \count 0 by 4
		\divide \count 4 by 4
	\else	\ifnum	\count 0 < -32767
		\then	\divide \count 0 by 4
			\divide \count 4 by 4
		\else
		\fi
	\fi
	\ifnum	\count 2 > 32767 
	\then	\divide \count 2 by 4
		\divide \count 4 by 4
	\else	\ifnum	\count 2 < -32767
		\then	\divide \count 2 by 4
			\divide \count 4 by 4
		\else
		\fi
	\fi
	\multiply \count 0 by \count 2
	\divide \count 0 by \count 4
	\xdef \product {#1 = \the \count 0 \internal@nits}%
	\aftergroup \product
       }}

\def\r@duce{\ifdim\dimen0 > 90\r@dian \then   
		\multiply\dimen0 by -1
		\advance\dimen0 by 180\r@dian
		\r@duce
	    \else \ifdim\dimen0 < -90\r@dian \then  
		\advance\dimen0 by 360\r@dian
		\r@duce
		\fi
	    \fi}

\def\Sine#1%
       {{%
	\dimen 0 = #1 \r@dian
	\r@duce
	\ifdim\dimen0 = -90\r@dian \then
	   \dimen4 = -1\r@dian
	   \c@mputefalse
	\fi
	\ifdim\dimen0 = 90\r@dian \then
	   \dimen4 = 1\r@dian
	   \c@mputefalse
	\fi
	\ifdim\dimen0 = 0\r@dian \then
	   \dimen4 = 0\r@dian
	   \c@mputefalse
	\fi
	\ifc@mpute \then
		\divide\dimen0 by 180
		\dimen0=3.141592654\dimen0
		\dimen 2 = 3.1415926535897963\r@dian 
		\divide\dimen 2 by 2 
		\Mess@ge {Sin: calculating Sin of \nodimen 0}%
		\count 0 = 1 
		\dimen 2 = 1 \r@dian 
		\dimen 4 = 0 \r@dian 
		\loop
			\ifnum	\dimen 2 = 0 
			\then	\stillc@nvergingfalse 
			\else	\stillc@nvergingtrue
			\fi
			\ifstillc@nverging 
			\then	\term {\count 0} {\dimen 0} {\dimen 2}%
				\advance \count 0 by 2
				\count 2 = \count 0
				\divide \count 2 by 2
				\ifodd	\count 2 
				\then	\advance \dimen 4 by \dimen 2
				\else	\advance \dimen 4 by -\dimen 2
				\fi
		\repeat
	\fi		
			\xdef \sine {\nodimen 4}%
       }}

\def\Cosine#1{\ifx\sine\UnDefined\edef\Savesine{\relax}\else
		             \edef\Savesine{\sine}\fi
	{\dimen0=#1\r@dian\advance\dimen0 by 90\r@dian
	 \Sine{\nodimen 0}
	 \xdef\cosine{\sine}
	 \xdef\sine{\Savesine}}}	      

\def\psdraft{
	\def\@psdraft{0}
}
\def\psfull{
	\def\@psdraft{100}
}

\psfull

\newif\if@scalefirst
\def\psscalefirst{\@scalefirsttrue}
\def\psrotatefirst{\@scalefirstfalse}
\psrotatefirst

\newif\if@draftbox
\def\psnodraftbox{
	\@draftboxfalse
}
\def\psdraftbox{
	\@draftboxtrue
}
\@draftboxtrue

\newif\if@prologfile
\newif\if@postlogfile
\def\pssilent{
	\@noisyfalse
}
\def\psnoisy{
	\@noisytrue
}
\psnoisy
\newif\if@bbllx
\newif\if@bblly
\newif\if@bburx
\newif\if@bbury
\newif\if@height
\newif\if@width
\newif\if@rheight
\newif\if@rwidth
\newif\if@angle
\newif\if@clip
\newif\if@verbose
\def\@p@@sclip#1{\@cliptrue}
\newif\if@decmpr
\def\@p@@sfigure#1{\def\@p@sfile{null}\def\@p@sbbfile{null}\@decmprfalse
   \openin1=\ps@predir#1
   \ifeof1
	\closein1
	\get@dir{#1}
	\ifx\ps@founddir\leer
		\openin1=\ps@predir#1.bb
		\ifeof1
			\closein1
			\get@dir{#1.bb}
			\ifx\ps@founddir\leer
				\ps@typeout{Can't find #1 in \figurepath}
			\else
				\@decmprtrue
				\def\@p@sfile{\ps@founddir\ps@dir#1}
				\def\@p@sbbfile{\ps@founddir\ps@dir#1.bb}
			\fi
		\else
			\closein1
			\@decmprtrue
			\def\@p@sfile{#1}
			\def\@p@sbbfile{#1.bb}
		\fi
	\else
		\def\@p@sfile{\ps@founddir\ps@dir#1}
		\def\@p@sbbfile{\ps@founddir\ps@dir#1}
	\fi
   \else
	\closein1
	\def\@p@sfile{#1}
	\def\@p@sbbfile{#1}
   \fi
}
\def\@p@@sfile#1{\@p@@sfigure{#1}}
\def\@p@@sbbllx#1{
		\@bbllxtrue
		\dimen100=#1
		\edef\@p@sbbllx{\number\dimen100}
}
\def\@p@@sbblly#1{
		\@bbllytrue
		\dimen100=#1
		\edef\@p@sbblly{\number\dimen100}
}
\def\@p@@sbburx#1{
		\@bburxtrue
		\dimen100=#1
		\edef\@p@sbburx{\number\dimen100}
}
\def\@p@@sbbury#1{
		\@bburytrue
		\dimen100=#1
		\edef\@p@sbbury{\number\dimen100}
}
\def\@p@@sheight#1{
		\@heighttrue
		\dimen100=#1
   		\edef\@p@sheight{\number\dimen100}
}
\def\@p@@swidth#1{
		\@widthtrue
		\dimen100=#1
		\edef\@p@swidth{\number\dimen100}
}
\def\@p@@srheight#1{
		\@rheighttrue
		\dimen100=#1
		\edef\@p@srheight{\number\dimen100}
}
\def\@p@@srwidth#1{
		\@rwidthtrue
		\dimen100=#1
		\edef\@p@srwidth{\number\dimen100}
}
\def\@p@@sangle#1{
		\@angletrue
		\edef\@p@sangle{#1} 
}
\def\@p@@ssilent#1{ 
		\@verbosefalse
}
\def\@p@@sprolog#1{\@prologfiletrue\def\@prologfileval{#1}}
\def\@p@@spostlog#1{\@postlogfiletrue\def\@postlogfileval{#1}}
\def\@cs@name#1{\csname #1\endcsname}
\def\@setparms#1=#2,{\@cs@name{@p@@s#1}{#2}}
\def\ps@init@parms{
		\@bbllxfalse \@bbllyfalse
		\@bburxfalse \@bburyfalse
		\@heightfalse \@widthfalse
		\@rheightfalse \@rwidthfalse
		\def\@p@sbbllx{}\def\@p@sbblly{}
		\def\@p@sbburx{}\def\@p@sbbury{}
		\def\@p@sheight{}\def\@p@swidth{}
		\def\@p@srheight{}\def\@p@srwidth{}
		\def\@p@sangle{0}
		\def\@p@sfile{} \def\@p@sbbfile{}
		\def\@p@scost{10}
		\def\@sc{}
		\@prologfilefalse
		\@postlogfilefalse
		\@clipfalse
		\if@noisy
			\@verbosetrue
		\else
			\@verbosefalse
		\fi
}
\def\parse@ps@parms#1{
	 	\@psdo\@psfiga:=#1\do
		   {\expandafter\@setparms\@psfiga,}}
\newif\ifno@bb
\def\bb@missing{
	\if@verbose{
		\ps@typeout{psfig: searching \@p@sbbfile \space  for bounding box}
	}\fi
	\no@bbtrue
	\epsf@getbb{\@p@sbbfile}
        \ifno@bb \else \bb@cull\epsf@llx\epsf@lly\epsf@urx\epsf@ury\fi
}	
\def\bb@cull#1#2#3#4{
	\dimen100=#1 bp\edef\@p@sbbllx{\number\dimen100}
	\dimen100=#2 bp\edef\@p@sbblly{\number\dimen100}
	\dimen100=#3 bp\edef\@p@sbburx{\number\dimen100}
	\dimen100=#4 bp\edef\@p@sbbury{\number\dimen100}
	\no@bbfalse
}
\newdimen\p@intvaluex
\newdimen\p@intvaluey
\def\rotate@#1#2{{\dimen0=#1 sp\dimen1=#2 sp
		  \global\p@intvaluex=\cosine\dimen0
		  \dimen3=\sine\dimen1
		  \global\advance\p@intvaluex by -\dimen3
		  \global\p@intvaluey=\sine\dimen0
		  \dimen3=\cosine\dimen1
		  \global\advance\p@intvaluey by \dimen3
		  }}
\def\compute@bb{
		\no@bbfalse
		\if@bbllx \else \no@bbtrue \fi
		\if@bblly \else \no@bbtrue \fi
		\if@bburx \else \no@bbtrue \fi
		\if@bbury \else \no@bbtrue \fi
		\ifno@bb \bb@missing \fi
		\ifno@bb \ps@typeout{FATAL ERROR: no bb supplied or found}
			\no-bb-error
		\fi
		\count203=\@p@sbburx
		\count204=\@p@sbbury
		\advance\count203 by -\@p@sbbllx
		\advance\count204 by -\@p@sbblly
		\edef\ps@bbw{\number\count203}
		\edef\ps@bbh{\number\count204}
		\if@angle 
			\Sine{\@p@sangle}\Cosine{\@p@sangle}
	        	{\dimen100=\maxdimen\xdef\r@p@sbbllx{\number\dimen100}
					    \xdef\r@p@sbblly{\number\dimen100}
			                    \xdef\r@p@sbburx{-\number\dimen100}
					    \xdef\r@p@sbbury{-\number\dimen100}}
                        \def\minmaxtest{
			   \ifnum\number\p@intvaluex<\r@p@sbbllx
			      \xdef\r@p@sbbllx{\number\p@intvaluex}\fi
			   \ifnum\number\p@intvaluex>\r@p@sbburx
			      \xdef\r@p@sbburx{\number\p@intvaluex}\fi
			   \ifnum\number\p@intvaluey<\r@p@sbblly
			      \xdef\r@p@sbblly{\number\p@intvaluey}\fi
			   \ifnum\number\p@intvaluey>\r@p@sbbury
			      \xdef\r@p@sbbury{\number\p@intvaluey}\fi
			   }
			\rotate@{\@p@sbbllx}{\@p@sbblly}
			\minmaxtest
			\rotate@{\@p@sbbllx}{\@p@sbbury}
			\minmaxtest
			\rotate@{\@p@sbburx}{\@p@sbblly}
			\minmaxtest
			\rotate@{\@p@sbburx}{\@p@sbbury}
			\minmaxtest
			\edef\@p@sbbllx{\r@p@sbbllx}\edef\@p@sbblly{\r@p@sbblly}
			\edef\@p@sbburx{\r@p@sbburx}\edef\@p@sbbury{\r@p@sbbury}
		\fi
		\count203=\@p@sbburx
		\count204=\@p@sbbury
		\advance\count203 by -\@p@sbbllx
		\advance\count204 by -\@p@sbblly
		\edef\@bbw{\number\count203}
		\edef\@bbh{\number\count204}
}
\def\in@hundreds#1#2#3{\count240=#2 \count241=#3
		     \count100=\count240	
		     \divide\count100 by \count241
		     \count101=\count100
		     \multiply\count101 by \count241
		     \advance\count240 by -\count101
		     \multiply\count240 by 10
		     \count101=\count240	
		     \divide\count101 by \count241
		     \count102=\count101
		     \multiply\count102 by \count241
		     \advance\count240 by -\count102
		     \multiply\count240 by 10
		     \count102=\count240	
		     \divide\count102 by \count241
		     \count200=#1\count205=0
		     \count201=\count200
			\multiply\count201 by \count100
		 	\advance\count205 by \count201
		     \count201=\count200
			\divide\count201 by 10
			\multiply\count201 by \count101
			\advance\count205 by \count201
		     \count201=\count200
			\divide\count201 by 100
			\multiply\count201 by \count102
			\advance\count205 by \count201
		     \edef\@result{\number\count205}
}
\def\compute@wfromh{
		\in@hundreds{\@p@sheight}{\@bbw}{\@bbh}
		\edef\@p@swidth{\@result}
}
\def\compute@hfromw{
	        \in@hundreds{\@p@swidth}{\@bbh}{\@bbw}
		\edef\@p@sheight{\@result}
}
\def\compute@handw{
		\if@height 
			\if@width
			\else
				\compute@wfromh
			\fi
		\else 
			\if@width
				\compute@hfromw
			\else
				\edef\@p@sheight{\@bbh}
				\edef\@p@swidth{\@bbw}
			\fi
		\fi
}
\def\compute@resv{
		\if@rheight \else \edef\@p@srheight{\@p@sheight} \fi
		\if@rwidth \else \edef\@p@srwidth{\@p@swidth} \fi
}
\def\compute@sizes{
	\compute@bb
	\if@scalefirst\if@angle
	\if@width
	   \in@hundreds{\@p@swidth}{\@bbw}{\ps@bbw}
	   \edef\@p@swidth{\@result}
	\fi
	\if@height
	   \in@hundreds{\@p@sheight}{\@bbh}{\ps@bbh}
	   \edef\@p@sheight{\@result}
	\fi
	\fi\fi
	\compute@handw
	\compute@resv}
\def\OzTeXSpecials{
	\special{empty.ps /@isp {true} def}
	\special{empty.ps \@p@swidth \space \@p@sheight \space
			\@p@sbbllx \space \@p@sbblly \space
			\@p@sbburx \space \@p@sbbury \space
			startTexFig \space }
	\if@clip{
		\if@verbose{
			\ps@typeout{(clip)}
		}\fi
		\special{empty.ps doclip \space }
	}\fi
	\if@angle{
		\if@verbose{
			\ps@typeout{(rotate)}
		}\fi
		\special {empty.ps \@p@sangle \space rotate \space} 
	}\fi
	\if@prologfile
	    \special{\@prologfileval \space } \fi
	\if@decmpr{
		\if@verbose{
			\ps@typeout{psfig: Compression not available
			in OzTeX version \space }
		}\fi
	}\else{
		\if@verbose{
			\ps@typeout{psfig: including \@p@sfile \space }
		}\fi
		\special{epsf=\ps@predir\@p@sfile \space }
	}\fi
	\if@postlogfile
	    \special{\@postlogfileval \space } \fi
	\special{empty.ps /@isp {false} def}
}
\def\DvipsSpecials{
	\special{ps::[begin] 	\@p@swidth \space \@p@sheight \space
			\@p@sbbllx \space \@p@sbblly \space
			\@p@sbburx \space \@p@sbbury \space
			startTexFig \space }
	\if@clip{
		\if@verbose{
			\ps@typeout{(clip)}
		}\fi
		\special{ps:: doclip \space }
	}\fi
	\if@angle
		\if@verbose{
			\ps@typeout{(clip)}
		}\fi
		\special {ps:: \@p@sangle \space rotate \space} 
	\fi
	\if@prologfile
	    \special{ps: plotfile \@prologfileval \space } \fi
	\if@decmpr{
		\if@verbose{
			\ps@typeout{psfig: including \@p@sfile.Z \space }
		}\fi
		\special{ps: plotfile "`zcat \@p@sfile.Z" \space }
	}\else{
		\if@verbose{
			\ps@typeout{psfig: including \@p@sfile \space }
		}\fi
		\special{ps: plotfile \@p@sfile \space }
	}\fi
	\if@postlogfile
	    \special{ps: plotfile \@postlogfileval \space } \fi
	\special{ps::[end] endTexFig \space }
}
\def\psfig#1{\vbox {
	\ps@init@parms
	\parse@ps@parms{#1}
	\compute@sizes
	\ifnum\@p@scost<\@psdraft{
		\PsfigSpecials 
		\vbox to \@p@srheight sp{
			\hbox to \@p@srwidth sp{
				\hss
			}
		\vss
		}
	}\else{
		\if@draftbox{		
			\hbox{\fbox{\vbox to \@p@srheight sp{
			\vss
			\hbox to \@p@srwidth sp{ \hss 
			 \hss }
			\vss
			}}}
		}\else{
			\vbox to \@p@srheight sp{
			\vss
			\hbox to \@p@srwidth sp{\hss}
			\vss
			}
		}\fi

	}\fi
}}
\psfigRestoreAt
\setDriver
\let\@=\LaTeXAtSign

\def \lta {\mathrel{\vcenter
          {\hbox{$<$}\nointerlineskip\hbox{$\sim$}}}} 
\def \gta {\mathrel{\vcenter
          {\hbox{$>$}\nointerlineskip\hbox{$\sim$}}}} 
\def\MPA#1#2{Max-Planck-Institut f\"ur Astrophysik 19{#1},
             Preprint {$\underline{#2}$} }
 \def\z{\phantom 1}
 \def\Hbar{$\overline H\ $}
 \def\Htil{$\widetilde H\ $}
 \def\Etilbv{$\widetilde E_{B-V}\ $}
 \def\etal{{et al.} \thinspace}
 \def\eg{{e.g.,} \thinspace}
 \def\ie{{i.e.,} \thinspace}
 \def\eck#1{\left\lbrack #1 \right\rbrack}
 \def\eqck#1{$\bigl\lbrack$ #1 $\bigr\rbrack$}
 \def\rund#1{\left( #1 \right)}
 \def\ave#1{\langle #1 \rangle}
 \def\:{\mskip\medmuskip}                         
 \def\lb{\lbrack} \def\rb{\rbrack}                
 \def\unit#1{\nobreak{\:{\rm#1}}}                 
 \def\inunits#1{\nobreak{\:\lb{\rm#1}\rb}}        
 \def\gcc{gcm$^{-3}$}
 \def\mstar{ M_{\ast} }
 \def\msol{ M_\odot }            
 \def\ms{ M_\odot }              
 \def\lsol{ L_\odot }            
 \def\ni{$^{56}{\rm Ni}\ $}      
 \def\Ni{$^{56}{\rm Ni}$}      
 \def\co{$^{56}{\rm Co}\ $}      
 \def\Co{$^{56}{\rm Co}$}      
 \def\fe{$^{56}{\rm Fe}\ $}      
 \def\kelvin{\thinspace\rm{\sp{o}{\kern-.08333em }K}\ }
\begin{document}
\title{Influence of the Stellar Population on Type Ia Supernovae:}
\title{Consequences for the Determination of $\Omega$ }
\bigskip
\bigskip
\centerline{P.~H\"oflich$^{1}$, K.Nomoto $^{2}$,H. Umeda $^{2}$, J.C. Wheeler $^{1}$}
\affil{1. Department of Astronomy, University of Texas, Austin, TX 078712, USA}
\affil{2. Department of Astronomy \& Research Center for the Early Universe, University of Tokyo,
 Tokyo 113-0033, Japan}

\begin{abstract}
 
The  influence of the metallicity at the main sequence 
on the chemical structure of the  exploding white dwarf, the nucleosynthesis during  the explosion
and the light curves  of an individual Type Ia  supernovae 
have been studied.
Detailed calculations of the stellar evolution, the explosion, and light curves
of delayed detonation models are presented.

 Detailed stellar evolution calculations with a main sequence mass 
$M_{MS}$ of 
7 $M_\odot $ have been performed to test the influence of the metallicity Z on the 
structure of the  progenitor.
A change of Z  influences the central helium burning and, consequently,
the size of the C/O core which becomes a C/O white dwarf
 and its C/O ratio. Subsequently,
the white dwarf may grow to the Chandrasekhar mass and explode as a Type Ia supernovae.
Consequently, the C/O structure of the exploding white dwarf depends on Z.
 Since C and O are the fuel for the thermonuclear explosion, Z indirectly changes the
energetics of the explosion. 
 
 In our example, changing Z from Pop I to Pop II causes 
a change in the zero point of the maximum brightness/decline relation by about $0.1^m$ and a change
in the rise time by about 1 day. Combined with  previous studies, 
the offset in the maximum brightness/decline $\Delta M \approx \Delta t $ 
 where $\Delta t$ is the change of the rise time in days.
 
Systematic effects of the size dissussed here may well make the results
from the SNe~Ia searches consistent with
an Universe with $\Omega_M=0.2$ and $\Omega_\Lambda=0$ but hardly
will change the conclusion that we live in a universe with low $\Omega_M$.
 Variations of the expected size may prove to be 
critical if, in the future, SNe~Ia are used to measure large scale scalar-fields because
 Z may show large local variations.
Evolutionary effects will not change substantially the counting rates for SNe~Ia even at very  large 
red-shifts.

Evolutionary effects may be of the same order  as the 
brightness changes related to cosmological parameters, but we have shown ways, how the effects 
of evolution can be detected.
 
\end{abstract}

\section{Introduction}

Two of the important new developments in observational supernova research
in the last few years were to establish the long-suspected
correlation between the peak brightness of SNe~Ia and their
rate of decline by means of modern CCD photometry (Phillips 1993) and the
exact distance calibrations provided by a  
 HST project (e.g. Saha et al. 1997). This allowed an empirical determination of $H_o$ with
unpreceded accuracy (Hamuy et al. 1996, Riess et al. 1995).
 Independent from these calibrations and empirical relations,
$H_o$ has been determined by   comparison of detailed 
theoretical models for light curves and spectra with observations 
(M\"uller \& H\"oflich 1994, H\"oflich \& Khokhlov 1996, Nugent et al. 1996). All determinations of the
Hubble constant are in good agreement with one another.
More recently, the routine successful detection of supernovae at 
large redshifts, z (e.g. Perlmutter et al. 1999,
 Riess et al. 1998), 
has provided an exciting new tool to probe cosmology.  This work has 
provided results that are
consistent with a low matter density in the Universe and, most
intriguing of all, yielded hints for a positive cosmological constant.  
 These results are based on empirical brightness-decline relations  which are
calibrated locally. This leaves systematic effects as the main source of concern.
 In this respect, theoretical models provide a critical tool  to
estimate the possible size of evolutionary effects, how these effects can be recognized and 
how one may be able  to  correct for them.

The primary scenario for SNIe~Ia consists of massive carbon-oxygen white
dwarfs (WDs) in a close binary system  which accrete
through Roche-lobe overflow from low mass companion star when it evolving  
away from the main sequence or during its red giant phase
(Nomoto \& Sugimoto 1977). The WD is the final evolution of stars
with a main sequence masses smaller than $\approx 8  ~M_\odot $ which has lost
his H/He rich envelope with the C/O core of $\approx 0.5 ... 1.2 M_\odot$.
 If accretion is sufficient large (i.e.
 $\geq 2 ... 4 ~10^{-8}M_\odot/yr$),  the accreted H burns  to He and, subsequently, to C/O 
on the surface of the WD and the mass of the WD grows close to the Chandrasekhar mass $M_{Ch}$.
 The corresponding time scales for growing to $M_{Ch}$ are $\leq   5\times  10^7 yrs$.
 In these accretion models, the explosion is
triggered by compressional heating close to the center.

H\"oflich, Khokhlov \& Wheeler (1995) showed that 
models based on Chandrasekhar mass carbon-oxygen white dwarfs 
can account for subluminous as wells as ``normally bright" SNe~Ia.
The basic paradigm of these models is that thermonuclear burning
begins as a subsonic, turbulent deflagration and then makes a 
transition to a supersonic, shock-driven detonation (Khokhlov 1991ab,  Yamaoka et al. 1992,  Woosley \& Weaver 1994).
These models are generally known as delayed detonation models.  
 In this class of models the amount of nickel 
produced is a function of the density at which the transition is 
made from deflagration to detonation, the central density, metallicity and  the chemical structure 
of the inital WD (H\"oflich et al. 1995, H\"oflich \& Khokhlov 1996, 
H\"oflich, Wheeler \& Thielemann 1998, HWT98 hereafter).
The radioactive decay of the variable nickel mass is the dominant factor which
gives a range in maximum brightness and $^{56}$Ni is the most important factor 
which governs the light curve shape.
The models with less nickel are not only dimmer, but are cooler 
and have lower opacity, giving them redder, more steeply declining 
light curves. A given   amount of nickel can be produced by different combinations 
of the model parameters and, from the models, we expect a spread around the mean 
maximum brightness-decline relation of $\approx ~0.3~$ to $0.5^m$.  
 A similar spread  of $\approx \pm 0.4^m$ ($\sigma \approx 0.18^m $ in B ) in the local, observed  
relation between maximum brightness  and decline (Hamuy et al. 1996) 
may also suggest that there  are additional parameters needed in the empirical relations.
 Note that new observations and  recalibrations of the old observations indicate
a somewhat tighter relation  ($\approx 0.12^m$, Jha et al. 1999; Kirshner et al., 1999). From the current status
of theoretical models, this very narrow spread cannot be understood, but it cannot be ruled out either.
 Both reanalysis of new light curves and theoretical investigation of the 
coupling between the ''free`` parameters may help to answer this puzzle. Note that this also implies that
observable spreads in relations, e.g. between rise time and decline time,
 cannot be expected to change along the path  given by the change of a particular free model parameter.

 There are  some hints that SN~Ia have undergone evolutionary effects.
 Branch et al. (1996) have  shown that the mean peak brightness is dimmer in ellipticals 
than in spiral galaxies. Wang, H\"oflich \& Wheeler (1997) found  that 
the peak brightness in the outer region of spirals is 
similar to those found in ellipticals, but that in the central region both 
intrinsically brighter and dimmer SNe~Ia occur.
  This implies that the underlying
progenitor populations are more inhomogeneous in the inner parts
of spiral galaxies which contain both young and old progenitors.
 Systematic effects that must be taken into account in the use of SNe~Ia to determine
cosmological parameters include technical problems, changes
of the environment with time,  changes in the statistical     
 properties of the SNe~Ia, and changes in the physical properties of SNe~Ia.

 In a previous paper (HWT98), the effect 
of the change of the initial metallicity and the C/O ratio on light curves and spectra
was studied by changing both the C/O ratio and the metallicity independently.
It was found that a reduction in the C/O ratio mainly effects the energetics of the explosion.
 The \ni production are reduced and the Si-rich layers are more confined in 
velocity space for smaller C/O.
 Moreover, the brightness to decline ratio can change causing a shift in the 
 zero point of the maximum brightness decline relation of $\approx 0.3^m$. Such a shift can 
be detected by a corresponding change in the rise time by about 3 days. 
 The influence of the metallicity is
especially important with respect to the $^{54}$Fe production in the outer layers which govern the
 spectra in the UV and blue; however, the influence on the resulting rest frame visual and blue light
curves is found to be small.
 A separation of the effects of Z and the C/O ratio in the WD is justified because 
 the metallicity is inherited from the ISM when it is formed and the C and O are produced 
during the progenitor evolution. To first order, the C/O ratio depends on the zero age main sequence mass.
Consequently, a systematic shift in the maximum brightness/decline relation can be produced by a
change with redshift of the typical progenitor mass. This effect is expected as the mean life time 
becomes shorter if we are looking back closer to the star forming period. 
 
 To second order, however, one must take into account that 
 the metallicity also influences the stellar evolution during  central helium burning.
The possible consequences for evolutionary effects in  SNIa and the
influence of $^{12}C(\alpha,\gamma)$-rate   as  has been pointed out
recently by various groups (Nomoto et al. 1997; Dominguez et al. 1999; 
H\"oflich et al. 1999; Umeda et al. 1999a).
 Thus, besides changing the C/O ratio as a consequence of the change in the typical progenitor mass, the effects
 of metallicity discussed above may also apply to individual SNe~Ia of a given main sequence mass.
 
 In this paper we study the latter effect: the influence of the metallicity on an individual
supernova with respect to the nucleosynthesis and light curves. We investigate  the size of the shift of the 
zero point in the brightness/decline relation  and
 the consequences for the use of SNIe~Ia for cosmology. We address the issue
how to distinguish the effect of a change in the ''typical`` progenitor from the effect 
of the metallicity
on the chemical structure of an individual progenitor.

\section{Brief Description of the Numerical Methods}
 
\subsection {Stellar Evolution}
 
The stellar evolution has been calculated using the code  of 
Nomoto's  group  up to the end of the helium burning. These calculations have already been published
and discussed in detail by  Umeda et al. (1999a).
 Subsequently, the evolution of the C/O core is calculated by 
accreting H/He rich material at  a given constant
 accretion rate on the core  by solving the standard equations for stellar evolution
using a Henye scheme.
 Nomoto's equation of state is used (Nomoto et al. 1982). Crystallization is neglected.
 For the energy transport, conduction (Itoh et al. 1983),
convection  in the mixing length theory, and radiation are taken into account. Radiative opacities for free-free and
bound-free transitions are treated in Kramer's approximation and by free electrons. A nuclear 
network of 35 species up to $^{24} Mg $ is taken into account.
 Here, we do not include convective mixing at the onset of 
 ignition of the deflagration front at the center. Depending on the 
description, the entire WD may mixed be mixed (Nomoto, Sugimoto \& Neo 
1976, H\"oflich et al. 1998). Although this may change the central 
carbon concentration at the center, the energetics of the explosions will
not change.

\subsection { Hydrodynamics}
 
The explosions are calculated using a one-dimensional radiation-hydro
code, including nuclear networks (H\"oflich \& Khokhlov 1996 and
references therein).
 This code solves the hydrodynamical equations
explicitly by the piecewise parabolic method (Collela \& Woodward 1984)
and includes the solution of the frequency averaged radiation transport
implicitly via moment equations, expansion opacities (see below),  and a detailed
equation of state.
 Nuclear burning is 
taken into account using a network which has been tested in many 
explosive environments (see Thielemann et al. 1996 and 
references therein).

\subsection { Light Curves}

Based on the explosion models, the subsequent expansion
 and  bolometric as well as broad band light
curves are calculated using a scheme recently developed, tested
and widely applied to  SN Ia (e.g. HWT98
and references therein).
The code used in this phase is similar to that described above, but
 nuclear burning is neglected and
 $\gamma $ ray transport is included via a Monte Carlo scheme.
In order to allow for a more consistent treatment of scattering, we
solve both the (two lowest) time-dependent, frequency averaged radiation moment equations for the
radiation
energy and the radiation flux, and a total energy equation.
At each time step, we then use $T(r)$ to determine the
Eddington factors and mean opacities by solving the frequency-dependent
radiation transport equation in the comoving frame 
and integrate to obtain the frequency-averaged quantities.
About one thousand frequencies (in one hundred frequency groups) and
about five hundred depth points are used. 
 The averaged  opacities
are calculated under the assumption
of local thermodynamical equilibrium. 
Both the monochromatic and mean opacities are calculated using the Sobolev
approximation
(Sobolev 1957).
The scattering, photon redistribution  and thermalization terms
used in the light curve opacity calculation are calibrated with NLTE
calculations using the formalism of the equivalent-two-level approach
(H\"oflich 1995).

\section{Results}

The initial mass of the C/O WD is given by the results of stellar evolution.
Its mass depends on  $M_{MS}$ of the progenitor and Z.
 At the time of the explosion, the WD masses are close to the Chandrasekhar limit. The WD has
grown by accretion of H/He and subsequent burning.
In the accreted layers, the C/O- ratio is close to 1.

\begin{figure}
\psfig{figure=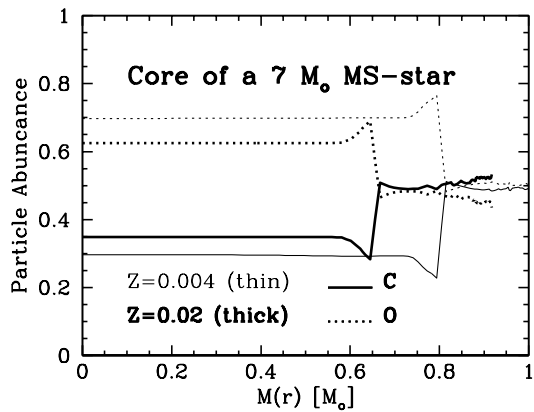,width=8.0cm,rwidth=8.0cm,clip=,angle=270}
\caption{ Chemical profile of the C/O core of a star with the main sequence mass of 
$ 7 M_\odot$ after  central Helium burning.}
\end{figure}

 Here, we study the size of the metallicity effect using  as an  example  a 7 $M_\odot $
with Population I (Z=0.02) and II (Z=0.004) compositions (Fig. 1). We assume that the relative abundance
within the metals is solar. 
In reality, the [O/Fe] changes with the metallicity and we may separate between $Z_{CNO}$ and $Z_{Fe}$.
The former governs the CNO cycle including the production of neon which determines the proton to 
nucleon ratio and, consequently, the abundances of the iron group elements produced during the thermonuclear
explosion. This will effect the spectra at maximum light (see HWT98) but hardly the size
of the C/O ratio which is the main subject of this letter. 
 It is the iron abundance that governs the opacity and changes the core size.
 The metallicity 
Z of heavy elements, e.g. Fe, mainly effects the convection during stellar
Helium burning and, consequently, the size of the C/O core and the central C/O ratio.
 We note that the exact size of the effect and its sign depends  on the mass of the progenitor
on the zero age main sequence, and Z. Even for a given mass, the changes are 
not monotonic,  but may change sign from Pop I to Pop II to Pop III (Umeda et al. 1999a,
Dominguez et al. 1999). In addition, the variation depends sensitively on the assumed physics such as the
$^{12}C(\alpha, \gamma )^{16}O$-rate (e.g. Straniero et al. 1997).
Our example can serve as a guide to estimate the size of this effect, but will not give all 
possible variations.

 At the time of the explosion, the central density has been fixed at $\rho_c=2.4E9 g/cm^{-3}$.
 The ratio  $\alpha$  of the deflagration velocity to the sound speed has been set to 0.02  and
 the transition density  $\rho_{tr}$ 
at which the  deflagration is assumed to turn into a detonation is
 is  $2.4~10^7 g~cm^{-3}$). The parameters  are close to those which
reproduce both the spectra and light curves reasonably well (Nomoto et al. 1984; H\"oflich
1995; H\"oflich \& Khokhlov 1996).
 
\begin{table*}
\caption{                          
Total abundances of the delayed detonation model for high and low metallicity}
\begin{tabular}{llllllllllll}
\hline
Population & He & C & O & Mg & Si & S & Ca & Fe  \\
 \hline
   I &    2.0E-04&     2.0E-02&     6.8E-02&     1.4E-02&     2.2E-01&     1.3E-01&     2.2E-02&     4.5E-01 \\  
 \hline
  II & 5.5E-05&     2.2E-02&     7.2E-02&     1.5E-02&     2.4E-01&     1.4E-01&     2.4E-02&     4.1E-01 \\  
 \hline           
\end{tabular}              
\end{table*}

\begin{figure}
\psfig{figure=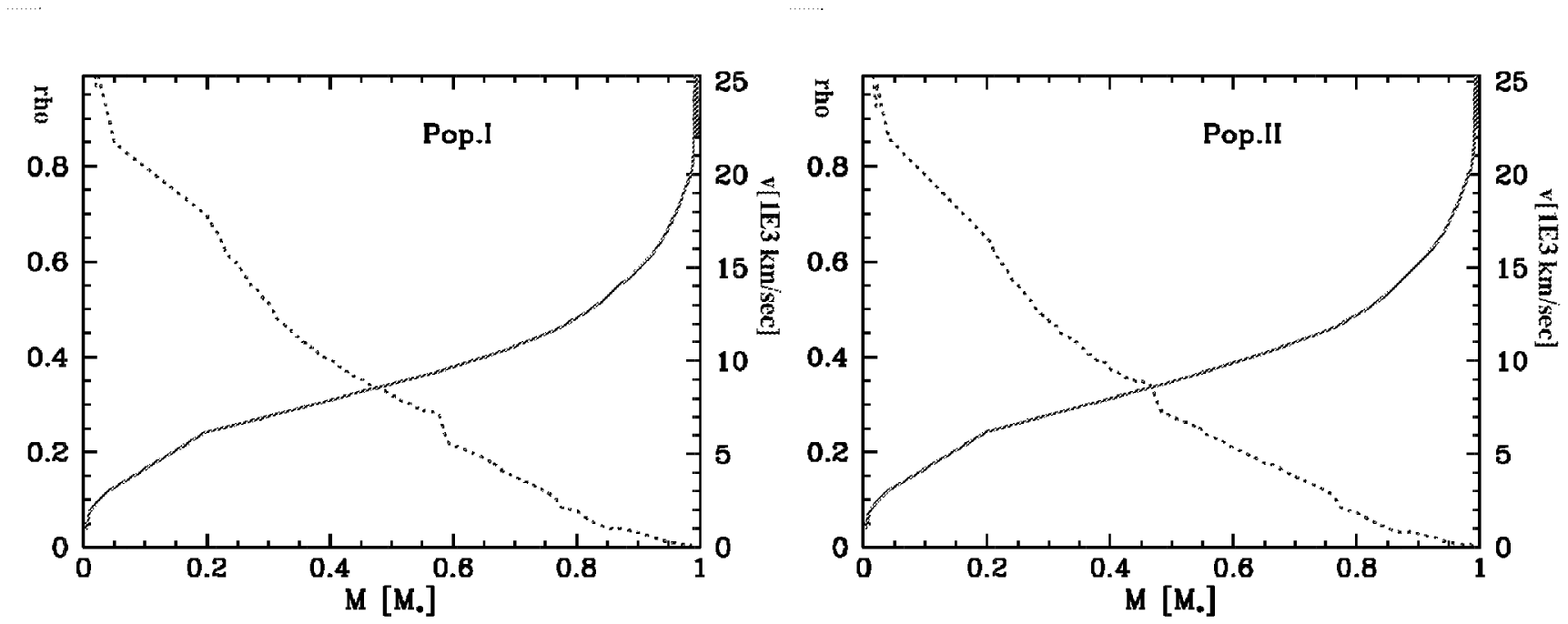,width=8.6cm,rwidth=8.5cm,clip=,angle=270}
\caption{ Density and velocity 
as a function of  mass for the delayed detonation model originating from 
progenitors with PopI and Pop II composition.
}
\end{figure}

\begin{figure}
 \psfig{figure=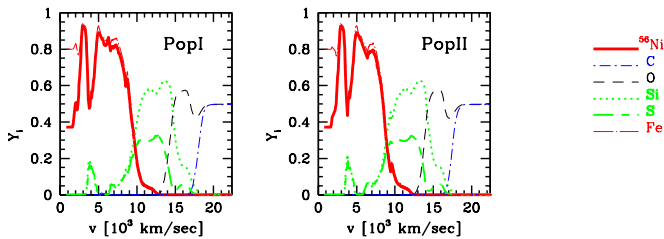,width=10.6cm,rwidth=10.5cm,clip=,angle=360}
\caption{ Same as Fig. 2, but abundances as a function of the final expansion velocity. 
Both the initial $^{56}$Ni and the final Fe profiles are shown.}
\end{figure}

 The total abundances of the most important elements are given for the  model  in 
Table 1. Fig. 2 gives the density and velocity versus mass for the models with the Pop I and Pop II metallicity,
respectively. Fig. 3 gives 
the composition profiles of the major elements for the same  models.
 The overall density and velocity structures are insensitive to changes in the underlying model.
Qualitatively, the final burning products can be understood by the relation between
hydrodynamic and the  nuclear time scales. 
Since most of the energy
is released by the explosive burning of carbon and oxygen, the energy production per gram 
depends on the initial chemical composition of the WD, i.e. the C/O ratio.
        The nuclear time 
scales are determined by the peak temperature during burning, which depends on the 
energy release per volume because the energy density is radiation dominated.
Since the energy release per gram is fixed, the peak temperature is given by the composition
and the local density.
 Thus, the latter variables are the dominant factors that determine the final composition of a zone.
After accretion on the initial core, the total mass ratio of $M_C/M_O$ is 0.75 and 0.61 for the 
models with Pop I and II,  respectively. The total  nuclear energy release depends on the 
total C/O ratio, but the structure of the WD is unaffected by a change in metallicity (HWT98).
Therefore, the final kinetic energy is reduced from $1.41x10^{51}erg$  for Z=0.02 to $1.37x10^{51}erg$ for Z=0.004.
The nuclear energy production during burning decreases with the C/O ratio at the burning front and, consequently,
the transition density is  reached later in time. The
resulting larger pre-expansion of the outer layers reduces the $^{56}$Ni production from 0.56  to
0.51 $M_\odot$ for  the Pop I and II model, respectively, and the transition regions between complete and incomplete
C and O burning changes by about 500 $km~s^{1}$.

\begin{figure}
\psfig{figure=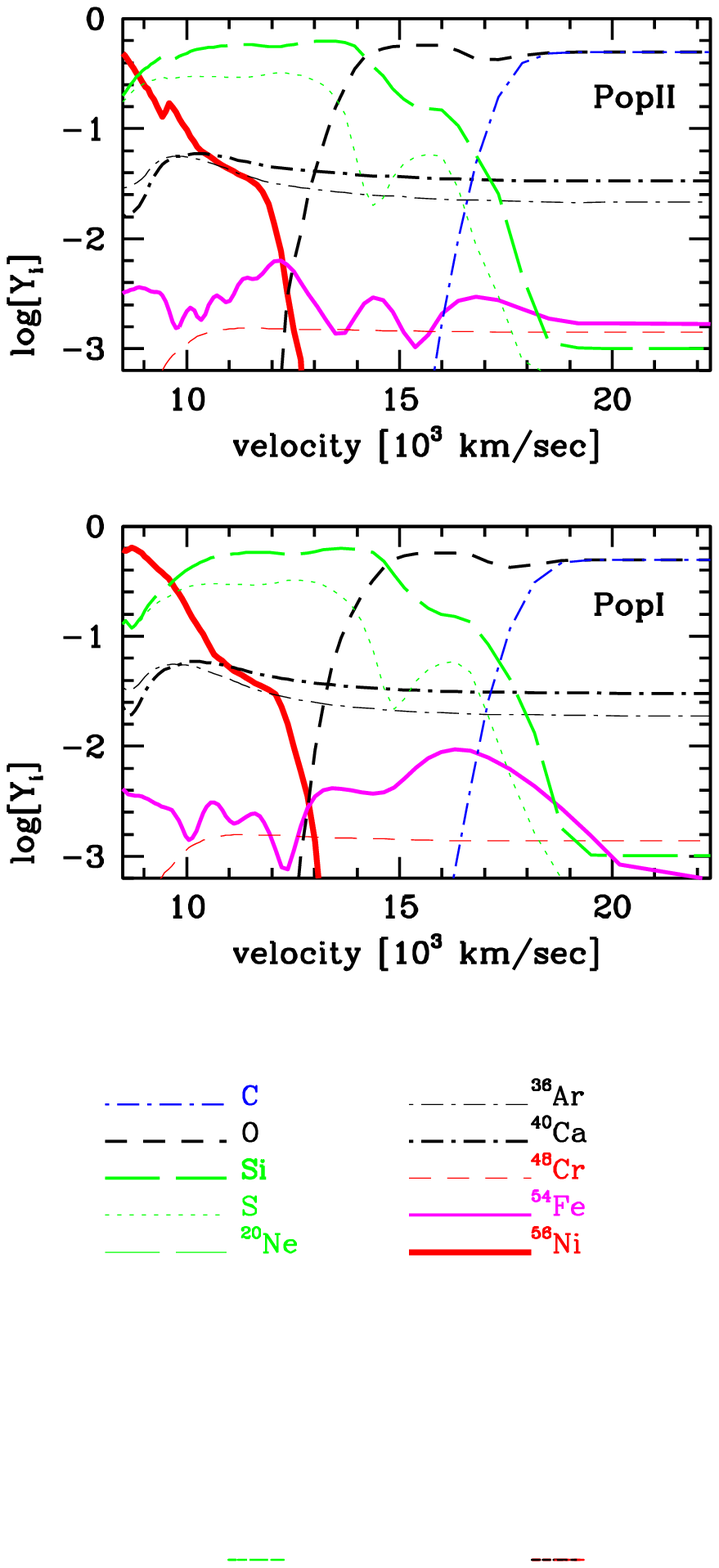,width=6.0cm,rwidth=6.0cm,clip=,angle=360}
\caption  {Same as Fig. 2, but abundances of different isotopes  as a function of the
expansion velocity for layers with partial burning.}
\end{figure}

 As discussed in HWT98, a change of the metallicity changes also the 
abundances of trace elements such as $^{54}$Fe (Fig. 4).
 The reason is that the metallicity mainly affects the initial CNO abundances of a star.
 These are converted
during the pre-explosion stellar evolution to $^{14}$N in H-burning and via 
$^{14}$N($\alpha,\gamma$) $^{18}$F($\beta ^+$) $^{18}$O($\alpha,\gamma$) $^{22}$Ne to nuclei with
N=Z+2 in He-burning. The result is that increasing metallicity yields
 a smaller proton to nucleon ratio, $Y_e$,                   
throughout the pre-explosive WD (Thielemann et al. 1997, HWT98). Higher metallicity and
 smaller $Y_e$
lead to the production of more neutron-rich Fe group nuclei and less $^{56}$Ni.
 For lower metallicity and thus
 higher $Y_e$, some additional \ni
is produced  at the expense of $^{54}$Fe  and $^{58}$Ni
  (Thielemann, Nomoto \& Yokoi, 1986).
 We note that these layers with v$\geq 12,000~km~ s^{-1}$ 
 dominate the spectra around maximum light and lines of the iron group, but have little influence on the light curves.

\begin{figure}
 \psfig{figure=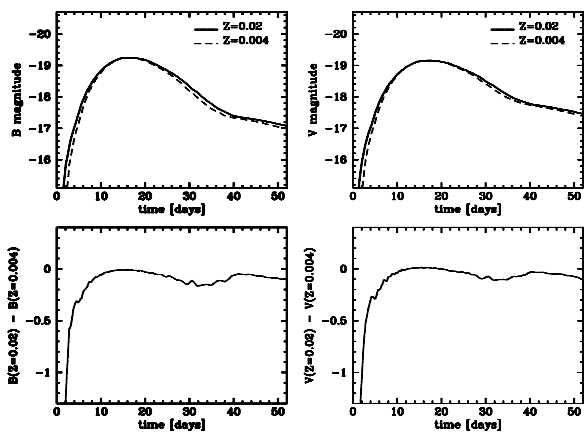,width=8.6cm,rwidth=8.5cm,clip=,angle=270}
\caption  {Comparison of  light curves in B  and V of the delayed detonation
models with metallicities corresponding to  Pop.I and Pop. II.}
\end{figure}

Light curves provide a valuable tool to probe the underlying 
explosion models, namely the absolute amount of \ni and its distribution.
Broad-band light curves are shown in Fig. 5. As expected from the last section,
 the main effect on the light curves is caused by the change in the explosion energy and the $^{56}$Ni
production.
 The change in the maximum brightness remains 
small ($M_V (Pop I) - M_V(Pop II)= -0.03^m$) and the rise times are different by about 1 day
($t_B((Pop I)=16.4d$ vs. $t_B(Pop II)=17.6 d$).
 The smaller expansion due to the smaller $E_{kin}$ causes a reduced geometrical dilution of the 
matter and a reduction of the expansion work at a given time.
 Consequently, for Pop II, the rise time
to maximum light is slower by about 1 days and the maximum brightness is only slightly smaller 
despite the 10 \% reduced production of
$^{56}$Ni   because, in the latter case, more of the
stored energy goes into the radiation rather than kinetic energy, and energy can be stored for a longer time.
 The most significant effect is a 
steeper decline ratio and a reduced $^{56}$Ni production for the Pop II-model.
 This translates into a systematic offset of $\approx 0.1^m$ in 
the maximum brightness decline ratio somewhat depending on the slope derive from the data
 (Phillips 1992, Hamuy et al. 1996). Using either the stretching method 
 or the LCS-method gives similar offsets. 
 
\section{Conclusions}
 
The  influence of the metallicity at the main sequence 
on the chemical structure of the  exploding white dwarf, the nucleosynthesis 
during  the explosion,
and the light curves  of  individual Type Ia  supernovae 
have been studied using  the example of delayed detonation models 
\footnote{Addendum: In this paper, we consider the dependence  of 
  the brightness-decline rate relation and changes in the rise
  times   on $X_C$ for given model parameters.
 We assume that the C/O ratio has little effect on the nuclear burning.
   In contrast, Umeda et al. (1999b)  considered the influence of $X_C$ on
 the transition density and, thus, on the production  of radioactive Ni.
  Fully consistent with previous papers (H\"oflich 1995, H\"oflich,
  Khokhlov \& Wheeler 1995), they find that the Ni production declines
  with the transition density between deflagration and detonation.
   The absolute brightness declines accordingly (HK96).
   Umeda et al. (1999) do not consider the effect of $X_C$ on the brightness-decline 
rate   relation or the rise times. Instead, for an assumed   relation between transition 
density and   C/O ratio, they discuss the implications on the typical brightness
  as a function of red-shift based on their study for progenitor evolution.}.
 Model parameters are used for which the
theoretical  optical  and infrared light curves  and of the spectral evolution resembles those of a 
normal bright SNe~Ia.  Note that we did not tune the parameter that the specific example resembles a particular
supernova or a specific normalization such as a stretch factor s=1.

A change of the metallicity influences the central helium burning and, consequently,
the size of the C/O core which becomes a C/O WD and its C/O ratio.
The WD
may grow to the Chandrasekhar mass and explode as a Type Ia supernovae.
Consequently, the metallicity changes the C/O structure of the exploding white dwarf.
 As C and O are the fuel for the thermonuclear explosion, the metallicity indirectly changes the
energetics of the explosion.  In addition, the metallicity  alters the isotopic composition 
of the outer layers of the ejecta that have undergone explosive O  burning.
 Especially important is the  increase of  the $^{54}$Fe production with metallicity which alters
the spectra  near maximum light (HWT98).

 As the C/O ratio of the WD is decreased, the explosion energy and 
 the \ni production are reduced.
 Changing the initial metallicity from Pop I to Pop II changes the
rise times by about 1 day, and causes a small decrease in luminosity at maximum light,
a  faster post-maximum decline and  a larger ratio between maximum light and \ni tail by about $0.1^m$.
This effect is equivalent to a change from a one-parameter maximum brightness/decline relation
which is widely used to determine the cosmological parameters $\Omega_M$ and $\Omega_\Lambda$. 
 In our example, the calibration of the $\Delta M(dM_{15})$ relation is changed by about $0.1^m$.
From previous studies of the stellar evolution for different masses and metallicites (
Umeda et al. 1999a, Dominguez et al. 1999),
 the influence of Z on the C/O ratio is comparable to our example, but the sign may
change depending on details of the nuclear cross sections and the treatment of convection, and the
mass and metallicity range.
 In our example, the systematic change in the progenitor metallicity from Pop I to Pop II translates into a shift of the confidence 
regions in the $\Omega_M-\Omega_\Lambda$ plane
based on high-z SNeI~a (Riess et al. 1998; Perlmutter et al. 1999)   by  about 0.15 and 0.7  
along the small and large axes of the error ellipsoids. 
However, a change of typical progenitor from Pop  I to II is  unlikely for the redshifts currently used to determine
$\Omega_M $ and $\Omega_\Lambda $ because we see a large variation of metallicities in our galaxy (Edvardsson et al.,
1993ab).
 
 Even strong variations in the metallicity will hardly produce selection effects with respect to the counting 
rates for SNe~Ia at large  red-shifts. Variations of the expected size
are  critical if, in future, SNe~Ia may be used to measure large scale scalar-fields because the metallicity
may show large local variations during the early phases when individual SNe explosions govern the
metallicity.
 
A change of the C/O ratio reveals itself mainly by the change
in the ratio  between rise time and decline. This change may be caused either by a change of Z or, alternatively,
by a change of the main sequence mass of the progenitor. 
When using statistical methods to determine cosmological parameters, both a change in the distribution of progenitor
masses and the metallicity of the sample may cause  systematic evolutionary effects.
 Fortunately, it does not matter whether a change of C/O is due to 
a change in the metallicity or the typical mass of the progenitors. In both cases, we find that
 $\Delta M \approx \Delta t_{rise} $ where $\Delta M $ is the offset in the maximum brightness/decline relation in magnitudes and
$\Delta t$  is measured in days (compare HWT98).
 
 The effect of progenitor mass and metallicity can be untangled 
by simultaneous analysis of both spectra and light-curves.  In principle, this allows a method 
to get a handle on the progenitor mass. 
With  the current level of modeling such  analysis is restricted to 
differential comparisons between individual
supernovae. 
 
 After submission of this paper, a comparison of rise times between local and distant supernovae has been
submitted (Riess et al. 1999). They  find a difference in the rise times between the local and the
distant sample of $2.5 \pm 0.4$ days if all observed supernovae are normalized to the stretch factor s of 1. 
This result is based on a preliminary analysis of the high redshift data by
 Goldhaber (1998). It may also be noted that both the local and the distant sample
show an  spread of $\pm 1 day$ in the rise time/decline relation without overlap which 
is hard to understand. However,
 if  confirmed and in light of our analysis, this would imply a systematic offset  $\approx 0.25^m$ in the 
brightness decline relation. A change of this order would not alter  the basic conclusion of the high
z searches that we live in a a low $\Omega$ universe but, then, 
a low $\Omega$ universe with  $\Omega_M=0.2$ and $\Omega_\Lambda=0$ may be consistent with  SN data as well.
 
The following trend for the theoretical models is worth noting:
For realistic cores, the both the mean $M_C/M_{O}$ and the $M_C/M_{O}$ in the central regions of the WD
 tend to  be smaller than the canonical value of 1
used in all calculations prior to 1998 (e.g. Nomoto et al. 1984, Woosley \& Weaver 1994, 
H\"oflich \& Khokhlov 1996).  Consequently, as a general trend, the rise times are about 
1-4 days slower compared to  models published before 1998.

 \subsection*{ACKNOWLEDGMENTS}
 
 We thank the referee for useful discussions and helpful comments.
This research was supported in part by  NASA Grant LSTA-98-022, NASA Grant NAG5-3930, NSF Grant AST-9528110,
and the grant-in-Aid for COE Scientific Research (07CE2002) of the
Ministry of Education, Science and Culture in Japan.
 The calculations for the explosion and light curves were done on a cluster of workstations 
financed by the John W. Cox-Fund  of the Department of Astronomy at the University of Texas.

{}
\end{document}